\newif\ifFIG
\FIGtrue

\newif\ifINCLUDE
\INCLUDEtrue

\newif\ifINCLUDENOT
\INCLUDENOTfalse

\documentclass[a4paper,11pt]{article} 

\usepackage{graphicx}

\usepackage{fullpage}
\usepackage{latexsym}
\usepackage{amsfonts}
\usepackage{amsmath}

\usepackage{subfig} 
\usepackage{epsf}
\usepackage{epstopdf}

\usepackage{xcolor}

\newcommand{\figurespath}{.}

\newcommand{\be}        { \begin{equation}  }
\newcommand{\ee}        { \end{equation}	}
\newcommand{\bea}       { \begin{eqnarray}  }
\newcommand{\eea}       { \end{eqnarray}    }

\begin{document}

\title{
\vspace{-0.3in}
\bf Who's the Daddy?\\
Infidelity Favours Father-Son Facial Dissimilarity
}

\author{
James V Stone, 
Psychology Department, Sheffield University, England.\\
Email: {\it j.v.stone@sheffield.ac.uk} \: 
File: whosthedaddy2021\_v22.tex. %
}

\maketitle

\vspace{-0.2in}
\noindent {\bf Abstract }
 Evolutionary theory predicts that children fare better if they resemble their father. However, if a man is promiscuous then   his children tend to be (unwittingly) raised within other families; such children should fare better if they do not resemble their father. This suggests that father-son facial similarity should decrease as a function of male promiscuity. To test this hypothesis, the degree of similarity between each of 37 pairs of father-son photographs was rated by 83 participants. 
The promiscuity score of a father was defined as the number of sexual relationships he had been involved in. 
A linear regression of father-son similarity ratings against fathers' promiscuity scores indicated that the slope of the fitted line was significantly less than zero (slope=-0.176, $p \textless 0.001$), and that fathers' promiscuity scores account for 75\% of the variance father-son facial similarity ratings. Additionally, a significant negative non-parametric correlation (Spearman's rho $\rho$=-0.886, $p$=0.019) between father-son similarity ratings and fathers' promiscuity scores was found. 


\section*{Introduction}
%
 Evolutionary theory predicts that children who resemble their father should attract more paternal investment than children who do not. For example, Christenfeld and Hill (1995) found that one year old children resembled their fathers, but not their mothers. However, the extent to which the prediction above is true depends on particular assumptions regarding the promiscuity of a child's father. Whereas a female always knows which children are hers, 
 a man cannot be certain that any of his mate's children were sired by him. Crucially, if a man suspects that he did not sire a specific child then that child is likely to receive unfavourable treatment.

The ultimate sanction for a child that does not resemble their father is infanticide, which may explain why infanticide is about 100 times as common in step-children as it is amongst children who live with their biological parents. For such children, the more they are perceived by their father to resemble him, the better they are treated in a population of fathers convicted of spouse abuse (Burch and Gallup, 2000). As might be expected, the tendency of perceived father-child similarity to engender favourable treatment also applies to more typical populations, where it has been found that father-child similarity is correlated with increased paternal investment (Apicella 2004; Alvergne et al, 2009; Platek, et al. 2003).

Thus, in the context of human evolution, the degree of father-child facial similarity should depend on how promiscuous the father is. At one extreme, if a man is monogamous then having children who look like him provides some protection for those children, because he can see which children were probably fathered by him. Conversely, a promiscuous man is likely to have children with women who already have a male partner. Such children would be easily detected by that male partner if they resemble their promiscuous father. This suggests that father-child similarity should decrease monotonically as the degree promiscuity of fathers increases. This line of reasoning is consistent with Pagel's mathematical model (Pagel, 2007), which predicts that father-child similarity is only advantageous to a child if his father is extremely monogamous.

\section*{Methods}
%
\noindent
{\bf Overview}. On each trial, a participant rated the similarity between a father-son pair of photographs. The mean similarity rating for different pairs was then plotted as a function of the fathers' promiscuity score (see Figure 1). 

The participants were 83 psychology undergraduates  (23 males and 60 females) from the University of Sheffield who participated for course credit (aged 18-29, mean = 19.3 years). Each pair of colour pictures depicted a male celebrity and his son. Each picture was 9cmx14cm, and the pictures in each father-son pair were presented side by side, separated by 9cm. Participants were approximately 57cm away from the screen. The 37 fathers were typically 40-50 years old, and the sons were typically 5-10 years of age. Faces contained no glasses or hats, and each face was oriented towards the camera. Faces were presented against a white background on a laptop; collars and other signs of clothing were removed digitally. 

\vspace{0.1in}
\noindent
{\bf Procedure}. Each participant was told they would see a series of father-son pairs of pictures. On each trial, the participant indicated the similarity between father and son by pressing a key from between 1 (not at all alike) and 9 (almost identical). Each pair of faces remained on the screen until the participant responded, when a 
fixation cross appeared for 0.5s, before another (randomly chosen) pair was presented. Each of the 37 father-son pairs was shown 3 times to each participant, making a total of 37x3=111 trials per participant. The experiment software was written using Psychopy (Peirce, 2007). 

\vspace{0.1in}
\noindent
{\bf Estimating Promiscuity}. Two measures of promiscuity were used: 
(a) the number N of women the father had relationships with, 
hereafter referred to as the promiscuity score. 
(b) the number of women C the father conceived a child with. 
Both promiscuity measures assigned to each father were based on a minimum of three reliable sources (e.g. broadsheet newspapers).

\section*{Results}
%
Promiscuity scores and face similarity scores are summarised in Table 1. 
A weighted linear regression indicates that father-son facial similarity ratings account for 75\% of the variance in mean promiscuity scores (r2=0.749, F=11.94, p=0.026), where each mean is the average similarity rating at one promiscuity score as shown in Figure 1; the fitted line has a slope significantly different from zero (slope=-0.176, sem=0.051, $p$ \textless 0.001). Because there is no a priori reason to assume that the relationship between similarity ratings and promiscuity scores is linear, we also calculated the non-parametric Spearman  correlation ($\rho$=-0.886, p=0.019). The other measure of promiscuity (C) provided no significant results, perhaps because these data have very little variance.

Because only one father-son pair had a promiscuity score of 10, this was excluded from all analyses. 
As a check, including this pair had a negligible effect on the conclusions reported here (e.g. if we include this pair then Spearman's correlation increased in magnitude to $\rho$=-0.893). 
All other data were included without any form of pre-processing. 

In order to check the robustness of these results, a bootstrap test (Efron, 1979) was performed on the similarity ratings\footnote{For each of 10,000 bootstrap samples, all responses of each participant (i.e. across picture-pairs) were re-sampled (with replacement) to obtain a re-sampled mean per father-son pair for that participant. These means were then combined across participants and father-son pairs to obtain a mean similarity rating for each promiscuity score (as in Figure 1), and the regression slope for these re-sampled similarity responses was estimated.}. This yielded a re-sampling distribution in which slopes equal to or less than the observed slope of -0.176 occurred with a statistically significant probability (p \textless 0.001). In other words, if each participant's observed similarity ratings were assigned at random to picture pairs then the probability of obtaining the results at least as extreme as those observed in the experiment is less than 0.001. 

To reflect the standard errors used in the conventional statistical analysis performed here, the standard errors in Table 1 and in Figure 1 were were computed after variance due to differences in participant means had been removed  (see Loftus and Masson, 1994); the data used for statistical analyses did not include any such normalisation.

\section*{Discussion}
A simplistic analysis of Darwinian theory suggests that children should resemble their fathers. However, a more detailed analysis predicts that the degree of father-child similarity should decrease as the promiscuity of fathers increases, and the results presented here are consistent with this prediction. 

Why should children resemble their fathers? Consider the following four possible scenarios. For clarity, we define a family as a husband and wife plus their children, and we define children sired by other men as step-children (and, for brevity, we state probabilistic tendencies as if they were definite facts). Within a family, on average:
\begin{enumerate}
\item	Monogamous husband and monogamous wife: Children of a monogamous husband and a monogamous wife should look more like the husband, because any step-children (e.g. resulting from rape) will be easier to spot by the husband, thus protecting children of the husband. Here, the mother and father's genes both benefit if the mother's genes are recessive and the father's genes are dominant with regard to facial appearance.
\item Monogamous husband and promiscuous wife: Children of a monogamous husband are protected if they resemble him, but the children of a promiscuous wife are protected if they resemble her. This way, step-children within the family will be harder to spot by the husband, thus providing some protection to the promiscuous wife's children sired by other men. Here, the mother and father's genes compete for dominance. 
\item	Promiscuous husband and monogamous wife: Children of a promiscuous husband are protected if they don't resemble him (i.e. they resemble their mother), and the children of a monogamous wife are protected if they resemble her husband. This way, his step-children in other families will be harder to spot than if his children resemble him, which therefore provides some protection to those step-children. Here, the mother and father's genes compete for recessiveness. 
\item	 Promiscuous husband and a promiscuous wife: Children of a promiscuous husband and a promiscuous wife should look more like the wife, for two reasons. First, her children will be sired by other men and by her husband, so her children sired by other men will be harder to spot by her husband, thus providing some some protection to her children. Second, his step-children (in other families) will be hard to spot by the husbands in those families if they look their mothers, thus providing some some protection to those children. Here, the mother and father's genes benefit if the mother's genes are dominant and the father's genes are recessive.
\end{enumerate}

\begin{table}[t]
\begin{center}
\begin{tabular}{|c|c|c|c|c|c|c|c|c|c|c|}
\hline
Promiscuity score, N & 1 & 2 & 3 & 4 & 5 & 6 & 7 & 8 & 9 & 10 \\
\hline
Number of father-son pairs & 15 & 10 & 3 & 3 & 3 & 2 & 0 & 0 & 0 & 1\\
\hline
Mean similarity score & 5.57 & 5.34 & 5.36 & 5.10 & 5.21 & 4.47 & - &  -&  - & 4.96\\
\hline
Standard error & 0.092    & 0.092    & 0.120    & 0.146    & 0.130   & 0.151 & - & - & -    & 0.217 \\
\hline
\end{tabular}
\end{center}
\caption{The promiscuity score (N) is the number of relationships a father had. 
The number of father-son pairs is the number of pairs in the study at each level of promiscuity.  
Each mean father-son similarity score was averaged over father-son pairs with the same promiscuity score. 
}
\label{default}
\end{table}%

\ifFIG
\begin{figure}[b!] %
\begin{center}
{\includegraphics[width=0.65\textwidth] {\figurespath/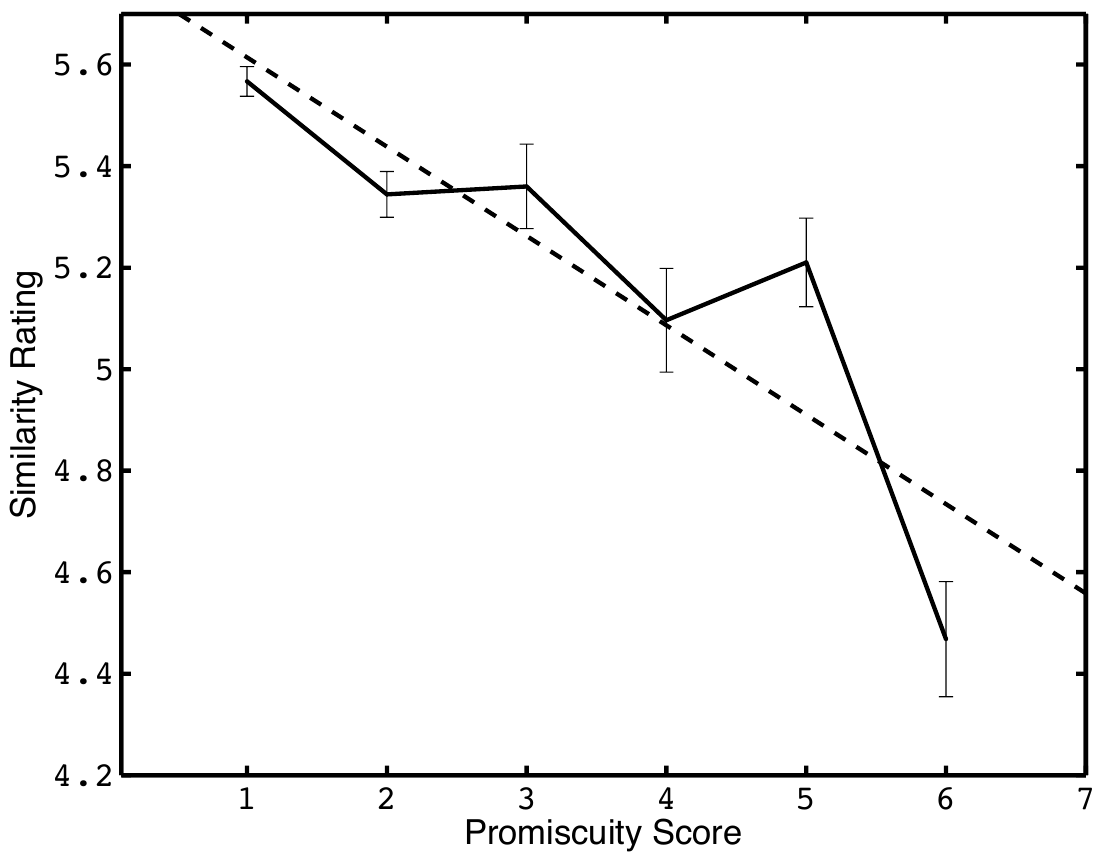} }
\caption{%
Father-son mean facial similarity ratings decrease with fathers' promiscuity scores, as shown by the best fitting line (dashed) obtained from a weighted regression analysis.  Bars are standard errors (see Results). 
}
\label{figpsych1}
\end{center}
\end{figure}
\fi

These arguments suggest that, if a child looks like its father then the father is likely to be monogamous (case 1), but the converse does not apply, because the presence of a monogamous father does not necessarily imply that the children in his family resemble him (due to case 2). However, overall, these four cases predict that father-child similarity should decrease as the promiscuity of fathers increases.

In practice, the prevalence of certain cases would naturally alter the strength of this prediction. However, it is noteworthy that only case 1 predicts strong father-child similarity. In the context of Pagel's mathematical model (Pagel, 2007), which concludes that ``even small amounts of paternity uncertainty can favour babies that adopt a strategy of anonymity'', the population would have to be dominated by extremely monogamous mothers and fathers in order for father-child similarity to be the norm. However, recent evidence suggests that women tend to be monogamous, whereas men tend to be promiscuous (Wlodarski et al, 2015), which also predicts that father-child similarity should decrease as the promiscuity of fathers increases. 

In this study, only father-son picture pairs were used. However, a natural extension would make use of father-daughter, mother-daughter and mother-son picture pairs. The current hypothesis predicts that father-child similarity should decrease with father promiscuity, but that mother-child similarity should increase with mother promiscuity.

Finally, it is noteworthy that the mean similarity ratings (for different promiscuity levels) chosen by participants varied between 4.5-5.6 (mean ratings for different father-son pairs varied between 4.0-7.5), even though participants could choose values in the range 1-9. Given that the father-son pairs used fathers and their actual sons, it is perhaps unsurprising that even the father-son pairs that were perceived as most dissimilar were not perceived as being very different (i.e. they had a mean similarity rating of about 4.4). Additionally, this may suggest that a smaller range of similarity ratings should have been made available to participants (e.g. between 0-3). Despite such considerations, the fact remains that 75\% of the variance in similarity ratings is accounted for by promiscuity scores.

\vspace{0.1in}
\noindent
{\bf Genetics}.  
We have not addressed the issue of how a mechanism which links promiscuity to facial similarity could arise.  
The epigenetic process known as genomic imprinting ensures that particular genes from one parent only are effectively silenced in an offspring. It is not hard to imagine that genomic imprinting is subject to competition between parents, which would ensure that  a child resembles either its father or mother. More generally, conventional gene dominance may give rise to one or other parent's genes being expressed more strongly in the offspring. If we assume that promiscuity is under genetic control (e.g. via testosterone) then the mechanism of gene linkage could account for an association between promiscuous behaviour and reduced father-child resemblance.
Whilst such an explanation is necessarily speculative, it should be borne in mind that the lack of a plausible mechanism has not always proved to be an insurmountable impediment to novel ideas within biology (e.g. Darwin, 1959)

\vspace{0.1in}
\noindent
{\bf Animal Studies.} Using a matching task, Rhesus monkeys performed better at identifying father-offspring than mother-offspring pairs, suggesting that offspring resemble their fathers more than their mothers. In contrast, chimpanzees performed better at mother-offspring than father-offspring pairs, suggesting that offspring resemble their mothers more than their fathers (Parr, 2010). However, whereas females are the dispersing gender in chimpanzees, males are the dispersing gender in Rhesus monkeys. As dispersing males begin life outside of their home group as low-ranking individuals, it is vital that their offspring are not recognised by high-ranking males. This, in turn, predicts that father-offspring matches should be easier to identify in chimpanzees than in Rhesus monkeys, as reported in (Parr, 2010). 

\vspace{0.1in}
\noindent
{\bf Confounds}.  
It is possible that participants had some familiarity with pictures of the fathers used in this experiment, and that this could have influenced the results. However, the stimuli (e.g. Frank Sinatra) used in this experiment were sufficiently old, and the participants (university students) were sufficiently young, that most fathers were not familiar to most of the participants. This was tested informally by asking a sample of participants if the fathers appeared to be familiar. However, given the surprisingly large size of the effect demonstrated here, a repeat of this experiment with more rigorous controls should be carried out.

\section*{Conclusion} 
Whereas a woman knows that all her children share her genes, a man can rarely be certain that any of his mate's children share his genes. For a species which seems to rely heavily on facial recognition, it would be surprising if a man did not discriminate against his mate's children according to how much they fail to  resemble him. In a population that contains even a small proportion of promiscuous men, such discrimination would place enormous selective pressure on children to resemble their father if he is monogamous, but to not resemble him if he is promiscuous. To our knowledge, the results presented are the first to provide support for this hypothesis. As Shakespeare wrote, ``it is a wise father that knows his own child'', but, for promiscuous men, it is a wise father that does not know his own child.




\vspace{0.1in}
\noindent
{\bf Acknowledgments}: Thanks to E Brookes, J Daniels, E Lucas, T Richardson, and N Ridley-Duff for preparing the stimuli and for testing participants.  Thanks to R Lister and two anonymous reviewers for their insightful comments and suggestions.

\newpage
\section*{References}
%
\bibliography{/Users/jimstone/Documents/biblio}
\bibliographystyle{apalike}

Alvergne, A, Faurie, C, and Raymond, M. (2009). Father?offspring resemblance predicts paternal investment in humans. Animal Behaviour, 78(1), 61-69.

\vspace{0.1in} \noindent
Apicella, CL and Marlowe, FW (2004). Perceived mate fidelity and paternal resemblance predict men's investment in children. Evolution and Human Behavior, 25(1), 371?378.

\vspace{0.1in} \noindent
Burch, RL, Gallup Jr, GG. (2000). Perceptions of paternal resemblance predict family violence. Evolution and Human Behavior, 21(6), 429-435.

\vspace{0.1in} \noindent
Christenfeld, N. J., \& Hill, E. A. (1995). Whose baby are you? Nature, 378(6558), 669-669.

\vspace{0.1in} \noindent
Darwin , C, (1859).The origin of species by means of natural selection.
 
 \vspace{0.1in} \noindent
Efron, B. (1979). Bootstrap methods: Another look at the jackknife. Ann. Statist., 7(1):1?26.

\vspace{0.1in} \noindent
Kazem, AJN and Widdig, A. (2013). Visual Phenotype Matching: Cues to Paternity Are Present in Rhesus Macaque Faces. PLoS ONE 8(2).

\vspace{0.1in} \noindent
Loftus, GR, and Masson, MEJ. (1994). Using confidence intervals in within-subject designs. Psychonomic Bulletin and Review, 1, 476-490.

\vspace{0.1in} \noindent
Pagel, M. (2007). Desperately Concealing Father: A Theory of Parent-Infant Resemblance, Animal Behaviour, 53(5), 973-981.

\vspace{0.1in} \noindent
Parr, LA, Heintz, M, Lonsdorf, E, Wroblewski, E. (2010). Visual kin recognition in nonhuman primates: Pan troglodytes and Macaca mulatta: inbreeding avoidance or male distinctiveness? Journal of Comparative Psychology 124:343?350.

\vspace{0.1in} \noindent
Peirce, JW. (2007). PsychoPy - Psychophysics software in Python. J Neurosci Methods, 162(1-2):8-13.

\vspace{0.1in} \noindent
Platek, S et al, (2003). 
How much paternal resemblance is enough? Sex differences in hypothetical investment decisions but not in the detection of resemblance. Evolution and Human Behavior 24, 81?87.

\vspace{0.1in} \noindent
Wlodarski, R, Manning, J, and Dunbar, RIM. (2015). Stay or stray? Evidence for alternative mating strategy phenotypes in both men and women. Biol. Lett. 11. 

\end{document}